\documentclass[twocolumn,showpacs,preprintnumbers,amsmath,amssymb]{revtex4}

\usepackage{graphicx}
\usepackage{dcolumn}
\usepackage{bm}

\input epsf

\begin{document}


\title{Thermopower of the Hubbard model: Effects of multiple orbitals and magnetic fields in the atomic limit}%

\author{Subroto Mukerjee}
\email{mukerjee@princeton.edu} 
\affiliation{Department of Physics, Princeton University}
\date{\today}
\begin{abstract}
We consider strongly-correlated systems described by the multi-orbital Hubbard model in the atomic limit and obtain exact expressions
for the chemical potential and thermopower. We show that these expressions reduce to the Heikes formula in the appropriate limits ($k_BT
\gg U$) and ($k_BT \ll U$) and obtain the full temperature dependence in between these regimes. We also investigate the effect of a
magnetic field introduced through a Zeeman term and observe that the thermopower of the multi-orbital Hubbard model displays spikes as
a function of magnetic field at certain special values of the field. This effect might be observable in experiments for materials with a
large magnetic coupling.

\end{abstract}

\maketitle

\section{\label{sec:intro}Introduction}

The thermopower of strongly correlated materials has received a lot of attention following the observation of a very large thermopower
at room temperature in NaCo$_{2}$O$_{4}$ \cite{terasaki}. These and other similar materials can be described by single or multi-orbital
Hubbard models with strong interactions. Several measurements of the thermopower of organic systems have been also carried out and the results 
were found to be describable by models of strongly interacting electrons on a lattice \cite{chaikine1, chaikine2}. In these models, the hopping 
energies were assumed to be much smaller than the correlation energies and temperatures. This limit is the so-called atomic 
limit. The effects of strong correlations on the thermopower has been studied in the whole range
described by the atomic limit earlier for the single band Hubbard model \cite{beni1}. The thermopower of the multi-orbital Hubbard model
has also been studied earlier, but only in the low and high temperature limits of the atomic limit \cite{chaikin, marsh}. In these
limits, the thermopower is essentially the entropy carried per unit charge and can be calculated from thermodynamics alone without using
the heat and charge currents of the system. A calculation over the entire atomic limit for the multi-band Hubbard model as done in this
paper requires explicit forms for the charge and heat currents which are derived later. Other approaches to the calculation of the
thermopower in these systems involve techniques such as dynamical mean-field theory \cite{palsson, merino} and exact diagonalization
\cite{koshibae1}.

Orbital degeneracy has a significant effect on the thermopower of a system. The increased entropy coming from the larger phase space due
to the orbitals serves to enhance the thermopower. Such an effect has indeed been observed in LaCrO$_{3}$ and LaMnO$_{3}$ where the
degeneracy of the 3$d$ orbitals causes an enhancement of thermopower \cite{raffaelle,stevenson}. Orbital degeneracy has also been
considered as a candidate to explain the high thermopower of the cobaltates \cite{koshibae2}. The presence of many orbitals however
brings up issues of Hund's rule coupling and energy level splitting which complicate the study of thermoelectric transport. These
effects are very difficult to take into account analytically if one is interested in calculating the temperature dependence of the
thermopower and have only been studied numerically \cite{koshibae2}. The Hubbard model without Hund's rule coupling does however lend
itself to analytic treatment in the atomic limit as will be demonstrated.

The effects of interaction on the thermopower beyond the atomic limit have been considered analytically for a single band Hubbard model
in certain cases. Oguri and Maekawa \cite{oguri} studied the $U=\infty$ model near half filling using a retraceable path approximation 
and Stafford \cite{stafford} used the Bethe ansatz to study the thermopower at low temperatures near half filling for the Hubbard chain. 
More recently, Paul and Kotliar \cite{paul} have calculated the thermopower in the Hubbard model in the infinite dimensional limit. The 
absence of an efficient large $U$ perturbation theory for the Hubbard model makes the calculation of the thermopower and other transport
coefficients difficult in the presence of strong correlations. The atomic limit is thus most amenable to transport calculations in
finite dimensions for a large range of temperatures and carrier densities.

A magnetic field is expected to suppress the thermopower by lifting the degeneracy of levels and reducing the entropy. Such an effect
has recently been observed experimentally in NaCo$_{2}$O$_{4}$ \cite{wang} where the magnetic field lifts the spin degeneracy of the
carriers drastically reducing the thermopower. The effect is expected to be more pronounced in the presence of orbital degeneracy. The
magnetic field affects not just the entropy but also that part of the thermopower which arises from a consideration of transport in the
system. It will be shown that one sees spikes in the thermopower at certain special values of the magnetic field when that effect is
considered.

In section II, we treat the single band Hubbard model in a magnetic field and get exact expressions for the chemical potential and 
thermopower. We show that the expression for the thermopower has particle-hole symmetry and reduces to that obtained by a consideration 
of free spins in the $U/k_BT \gg 1$ limit. Section III deals with the multiple orbital Hubbard model. Here we consider a system with 
degenerate orbitals and no Hund's rule coupling and derive expressions for the thermopower and fugacity as a function of temperature. We 
obtain the appropriate particle-hole symmetric Heikes expressions by considering the two limits $U/k_BT \gg 1$ and $U/k_BT \ll 1$. 
Finally, in section IV, we consider the multi-orbital Hubbard model in the presence of a magnetic field. We notice the appearance 
of spikes for certain special values of the magnetic field and demonstrate how this follows as a consequence of energy conservation.

\section{\label{sec:magsingle}Single orbital Hubbard model in a magnetic field}

The Hamiltonian we wish to consider is
\begin{equation}
H = -t \sum_{j \sigma}c_{j \sigma}^+c_{j+1 \sigma} + {\rm h.c.} + U \sum_{j}n_{j \uparrow}n_{j 
\downarrow} + gB\sum_{j}(n_{j \uparrow} - n_{j \downarrow})
\label{singmag}
\end{equation}
where $t$ is the hopping parameter, $U$ the interaction, $g$ the Zeeman coupling and $B$, the magnetic field. The index $j$ runs over
all the sites in the chain. The Hamiltonian as written corresponds to a single orbital Hubbard chain with a magnetic field that couples
to the carriers through only a Zeeman term. We note that all calculations in the atomic limit are carried out to lowest order in the
hopping $t$ (order $t^2$) and to this order, the results are the same for higher dimensional lattices as well. We also note that for
higher dimensional lattices, there is a term in the Hamiltonian arising from the magnetic field that couples to the hopping
parameter. This term can also be neglected to lowest order in perturbation theory in the atomic limit. The expressions for the charge
and heat current operators $J_e$ and $J_Q$ are given by
\begin{eqnarray}
J_e & = & \lim_{k \rightarrow 0}\frac{q}{\hbar k}\sum_{j}[n_j, H]e^{ikja} \\
J_Q & = & \lim_{k \rightarrow 0}\frac{1}{\hbar k}\sum_{j}[h_j, H]e^{ikja} \nonumber
\label{currents}
\end{eqnarray}
where $n_j$ and $h_j$ are the local charge and energy densities, $a$, the lattice constant and 
$q$, the charge of a carrier.
For the Hamiltonian given by Eqn. \ref{singmag}, the currents are
\begin{equation}
J_e = \frac{qat}{i\hbar}\sum_{j \sigma}c_{j \sigma}^+c_{j-1 \sigma}-c_{j \sigma}^+c_{j+1 \sigma}
\label{singe}
\end{equation}
\begin{eqnarray}
J_Q & = & \frac{at}{i\hbar}\sum_{j \sigma}(c_{j \sigma}^+c_{j-1 \sigma}-c_{j-1 \sigma}^+c_{j \sigma})(Un_{j -\sigma}+ g_\sigma B) 
\nonumber \\ 
& & + \frac{at^2}{i\hbar}\sum_{j \sigma}c_{j \sigma}^+c_{j-2 \sigma}-c_{j-2 \sigma}^+c_{j \sigma}
\label{singQ}
\end{eqnarray}
Here $g_\uparrow = g$ and $g_\downarrow = -g$.
The thermopower is given by
\begin{equation}
S=-\frac{L_2/L_1 + \mu/q}{T}
\label{thermo}
\end{equation}
where $T$ is the temperature and $\mu$ the chemical potential. The coefficients $L_2$ and $L_1$ 
are proportional to the Peltier coefficient and conductivity respectively and are given by the 
Kubo formulae.
\begin{equation}
L_1 = \frac{\int_{0}^{\infty}d\tau {\rm Tr}[e^{-\beta(H-\mu N)}(J_eJ_e(\tau)+J_eJ_e(\tau))]}{{\rm 
Tr}[e^{-\beta(H-\mu N)}]}
\label{kubo1}
\end{equation} 
\begin{equation}
L_2 = \frac{\int_{0}^{\infty}d\tau {\rm Tr}[e^{-\beta(H-\mu N)}(J_QJ_e(\tau)+J_eJ_Q(\tau))]}{{\rm
Tr}[e^{-\beta(H-\mu N)}]}
\label{kubo2}
\end{equation} 
$N$ is the number of carriers and $\beta = 1/k_{B}T$. The time shifted current operators are given 
by 
\begin{equation}
J_\alpha(\tau)= e^{\tau H} J_\alpha e^{-\tau H}
\end{equation}
It should be noted that both $L_1$ and $L_2$ are infinite in the atomic limit but their ratio is not. This is because the decays of the
current-current correlation functions are assumed to be the same in both the heat and charge channels. This assumption presumably breaks
down in the limit where real dissipation is introduced at higher orders of $t/U$. The procedure outlined from here on is similar to the
calculation by Beni \cite{beni1} in the absence of a magnetic field. We calculate the coefficients $L_1$ and $L_2$ to lowest order in
$t$ (order $t^2$) and keep the hopping parameter in the calculation only in the current and not the exponential factors. We can also
neglect the second term in the expression for the heat current Eqn. \ref{singQ}.  The nearest neighbor hopping allows us to take the
trace over only a pair of nearest neighbor sites (1 and 2) with the constraint of energy conservation $n_{2 -\sigma} = n_{1 -\sigma}$.
The resultant expression for the thermopower is
\begin{equation}
S=-\frac{k_B}{q}\left[\frac{\beta Ux^2}{e^{\beta U}+x^2} + \beta gB \tanh(\beta gB)\left(\frac{x^2-e^{\beta U}}{x^2+e^{\beta U}} 
\right) - \ln x \right]
\label{singS}
\end{equation}
$x=e^{\beta \mu}$ is the fugacity and can be obtained from the equation 
\begin{equation}
\frac{\partial Z}{\partial x} = \rho \frac{Z}{x}
\label{fugacity}
\end{equation}
where $Z = {\rm Tr}e^{-\beta(H-\mu N)}$ is the partition function and $\rho$ is the carrier density, with half-filling corresponding to 
$\rho$=1. In the present case, $x$ is the root of a quadratic equation and is given by 
\begin{widetext}
\begin{equation}
x = \frac{(\rho-1)\cosh(\beta gB) +\sqrt{(1-\rho)^2\cosh^2(\beta gB) + \rho(2-\rho)e^{-\beta U}}}{(2-\rho)e^{-\beta U}} 
\end{equation}
\end{widetext}
The first two terms of the expression for the thermopower Eqn \ref{singS} correspond to the contribution form transport and the 
third term corresponds to the entropy. The temperature dependence is given by the first two terms which disappear in the Heikes 
limit. In the limit of large $\frac{U}{k_{B}T}$, the thermopower reduces to
\begin{widetext}
\begin{equation}
S = \frac{k_B}{q}[\underbrace{\ln \rho - \ln (1-\rho)}_{\rm configurational} - \underbrace{\ln\{2\cosh(\beta gB)\} + \beta gB\tanh(\beta 
gB)}_{\rm spin}]
\end{equation} 
for $\rho < 1$ and
\begin{equation}
S = \frac{k_B}{q}[\underbrace{\ln (\rho-1) - \ln (2-\rho)}_{\rm configurational} + \underbrace{\ln\{2\cosh(\beta gB)\} - \beta 
gB\tanh(\beta gB)}_{\rm spin}]
\end{equation}
for $\rho > 1$.
\end{widetext}
where the first term corresponds to the configurational entropy and the second term gives the spin entropy corresponding to free spins. 
The full temperature dependence is shown in Fig. 1. We can see that the thermopower is particle-hole symmetric and disappears at half
filling ($\rho = 1$). For low and high values of the carrier density, the thermopower goes down in magnitude with increasing magnetic 
field (or decreasing temperature) corresponding to the reduction of entropy. For intermediate values, the thermopower changes sign going 
from particle like to hole like or vice versa because of the effect of the transport term. Indeed, the transport term can cause a 
change in the sign of the thermopower as a function of temperature even in the absence of a magnetic field \cite{beni1}. This effect 
is suppressed as one approaches the Heikes limit.
\begin{figure}
\label{singlemag} 
\epsfxsize=3in 
\centerline{\epsfbox{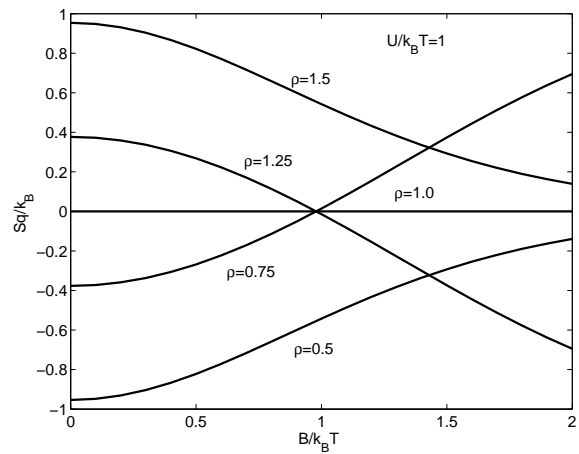}}
\caption{Thermopower as function of magnetic field for the single band Hubbard model. The curves are symmetric about half filling $\rho 
= 1$ and the thermopower goes down with applied field for low and high carrier concentrations. For intermediate values, the thermopower 
changes sign with field.} 
\end{figure}

\section{\label{sec:multi}Multiple orbital Hubbard model}
Let us now look at the multiple orbital Hubbard model. We assume that every site has a set of degenerate levels for the carriers and
there is a probability for a carrier to hop from a level on one site to any level on an adjacent site. In addition, the carriers
interact with all other carriers on the same site with an interaction strength $U$. A more general model would include different
interaction strengths for each set of orbitals and exchange energies (Hund's rule couplings). However such a model is not easy to treat 
analytically even at the atomic level and we consider this simplified model which still captures a lot of the essential physics. Our 
Hamiltonian is thus
\begin{eqnarray}
H & = & -\sum_{j\alpha \beta \sigma}t_{\alpha \beta}c_{j\alpha \sigma}^+c_{j+1\beta \sigma}
+ {\rm h.c.} + U \sum_{j\alpha}n_{j\alpha \uparrow}n_{j\alpha \downarrow} \nonumber \\ & & + \frac{U}{2}\sum_{\stackrel {j\alpha \beta
\sigma \sigma'}{\alpha \neq \beta}}n_{j\alpha \sigma}n_{j\beta \sigma'} 
\label{multib}
\end{eqnarray}
The currents are once again given by Eqn. \ref{currents} and in this case are
\begin{equation}
J_e = \frac{qa}{i\hbar}\sum_{j\alpha \beta \sigma}t_{\alpha \beta}c_{j+1\alpha \sigma}^+c_{j\beta \sigma}-t_{\alpha \beta}^{*}c_{j\alpha 
\sigma}^+c_{j+1\beta \sigma} 
\label{multie}
\end{equation}
\begin{widetext}
\begin{eqnarray}
J_Q & = & \frac{Ua}{i\hbar}\left[\sum_{j\alpha \beta \sigma}(t_{\alpha \beta}c_{j+1\alpha \sigma}^+c_{j\beta \sigma}
-t_{\alpha \beta}^*c_{j\beta \sigma}^+c_{j+1\alpha \sigma})n_{j+1\alpha -\sigma} +  
\frac{1}{2}\sum_{\stackrel{j\alpha \beta \gamma \sigma \sigma'}{\gamma \neq \alpha}}(t_{\gamma \beta}c_{j+1\gamma
\sigma'}^+c_{j\beta \sigma'}-t_{\gamma \beta}^*c_{j\beta \sigma'}^+c_{j+1\gamma \sigma'})n_{j+1\alpha \sigma} \right] \nonumber \\
& & + \frac{a}{i\hbar}\sum_{j\alpha \beta \gamma \sigma}t_{\alpha \beta}t_{\beta \gamma}c_{j+2 \gamma \sigma}^+c_{ja \sigma}-t_{\alpha
\beta}^*t_{\beta \gamma}^*c_{j \alpha \sigma}^+c_{j+2 \gamma \sigma}
\label{multiQ}
\end{eqnarray}
\end{widetext}
Here $\alpha$, $\beta$ and $\gamma$ are orbital indices while $\sigma$ and $\sigma'$ are spin indices. 
The Kubo formulae Eqns. \ref{kubo1} and \ref{kubo2} give the following upon reduction of the problem to a trace over two sites as 
before. 
\begin{equation}
\frac{L_2}{L_1} = \frac{I_2}{I_1}
\label{trace}
\end{equation}
where
\begin{equation}
I_1 = {\rm Tr} [B(\{n_{1\alpha \sigma_1},n_{2\beta \sigma_2}\})\sum_{\alpha \beta \sigma}|t_{\alpha \beta}|^2(n_{2\alpha 
\sigma}-n_{1\beta \sigma})^2] 
\end{equation}
and
\begin{widetext}
\begin{equation}
I_2  = {\rm Tr} [B(\{n_{1\alpha \sigma_1},n_{2\beta \sigma_2}\})U\{\sum_{\alpha \beta \sigma}|t_{\alpha \beta}|^2(n_{2\alpha
\sigma}-n_{1\beta \sigma})^2 n_{2\alpha -\sigma} + \sum_{\alpha \beta \gamma \sigma \sigma'}|t_{\gamma \beta}|^2(n_{2\gamma \sigma'}
-n_{1\beta \sigma'})^2 n_{2\alpha \sigma}\}]
\end{equation}
and the traces have to be performed subject to the constraint that 
\begin{equation}
U(n_{1\beta -\sigma} + \sum_{\gamma \neq \beta \sigma'}n_{1\gamma \sigma'} - n_{2\alpha -\sigma} - \sum_{\gamma \neq \alpha 
\sigma'}n_{2\gamma \sigma'}) = 0
\label{constr1}
\end{equation}
\end{widetext}
which is just the condition of energy conservation that says that the number of carriers on any two sites participating in a hop should 
be the same. We have used the fact that $t_{\alpha \beta} = t_{\beta \alpha}^*$ in the above expressions. In order to perform the traces, 
we make a further simplifying assumption that $t_{\alpha \beta} = t$ for all $\alpha$ and $\beta$. With this simplification, we are able 
to obtain the exact expression
\begin{equation}
S = \frac{k_B}{q}\left(\frac{U\sum_{n=0}^{2N-1}n{2N-1 \choose n}^2x^{2n+1}e^{-\beta n^2 U}}{\sum_{n=0}^{2N-1}{2N-1 \choose 
n}^2x^{2n+1}e^{-\beta n^2 U}} - \ln x \right)
\label{multiST}
\end{equation} 
where $N$ is the number of orbitals per site and the fugacity $x$ is given from Eqn. \ref{fugacity} by
\begin{equation}
\sum_{n=0}^{2N}{2N \choose n}(n - \rho)e^{-\beta U n(n-1)/2}x^n = 0
\end{equation}
It is straightforward to show that the above equation has exactly one non-negative root for all values of $\rho$. For $U \gg k_B T$, 
we obtain 
\begin{equation}
x=\frac{\rho}{2N(1-\rho)}
\end{equation} 
which is positive only for $\rho < 1$ thereby preventing double occupancy. This gives us the multi-orbital Heikes formula
\begin{equation}
S = \frac{k_B}{q}[\underbrace{\ln \rho - \ln (1-\rho)}_{\rm configurational} - \underbrace{\ln 2}_{\rm spin} - \underbrace{\ln N}_{\rm 
orbital}]
\end{equation}
It should be noted that Eqn. \ref{multiST} is particle-hole symmetric about half filling $\rho = N$. For $\rho < N$, the transport
part ($L_2/L_1$) goes to zero in the $U/k_BT \gg 1$ limit. For $\rho > N$, however, the contribution of the transport part has to be
considered in addition to the entropic part and only a proper addition gives the right Heikes limit. For $\rho > N$, we 
obtain
\begin{equation}
S = -\frac{k_B}{q}[\underbrace{\ln (2N-\rho) - \ln (\rho-2N+1)}_{\rm configurational} - \underbrace{\ln 2}_{\rm spin} - \underbrace{\ln 
N}_{\rm orbital}]
\end{equation}
$S$ is thus finite only for $\rho \leq 2N-1$ if $\rho > N$. For values of $\rho$ between 1 and $2N-1$, the thermopower is divergent in 
this limit except at half filling ($\rho = N$) where it is identically zero. 
In the opposite limit $U \ll k_B T$, one obtains
\begin{equation}
x=\frac{\rho}{2N-\rho}
\end{equation}
The thermopower in this limit is
\begin{equation}
S=-\frac{k_B}{q}\ln \frac{2N-\rho}{\rho}
\label{gheikes}
\end{equation}
where all values of $\rho$ between 0 and $2N$ give a finite thermopower. Eqn. \ref{gheikes} is the so-called generalized Heikes formula
previously obtained through direct combinatrics \cite{chaikin,koshibae1}. The expression for the thermopower in this limit is explicitly
particle-hole symmetric since the contribution to the thermopower for all values of $\rho$ comes only from the entropic term. The
temperature dependence between the two limits considered above is given by Eqn. \ref{multiST}. The thermopower changes sign as a
function of filling both below and above half-filling. Fig.2 shows the dependence of thermopower on temperature and carrier
concentration for $N=3$. Below half filling, the thermopower changes sign at $\rho=2N/(2N+1)$ which for $N=3$ is $\rho=6/7$, for small
$k_BT/U$ and always remains negative for small $k_BT/U$. Thus the curves with $\rho > 6/7$ change sign as a function of the parameter
$k_BT/U$. It can also be seen that the curves are not monotonic at a fixed value of $k_BT/U$ with $\rho/2N$ and display oscillations.  
The number of maxima on either side of half-filling ($\rho/2N = 0.5$) is equal to the number of orbitals. The oscillations disappear in 
the limit of large $k_BT/U$ and all the curves collapse onto a smooth curve as given by Eqn. \ref{gheikes}.
\begin{widetext}
\begin{figure}
$\begin{array}{cc}
\epsfxsize=3in
\epsfysize=3in
\epsffile{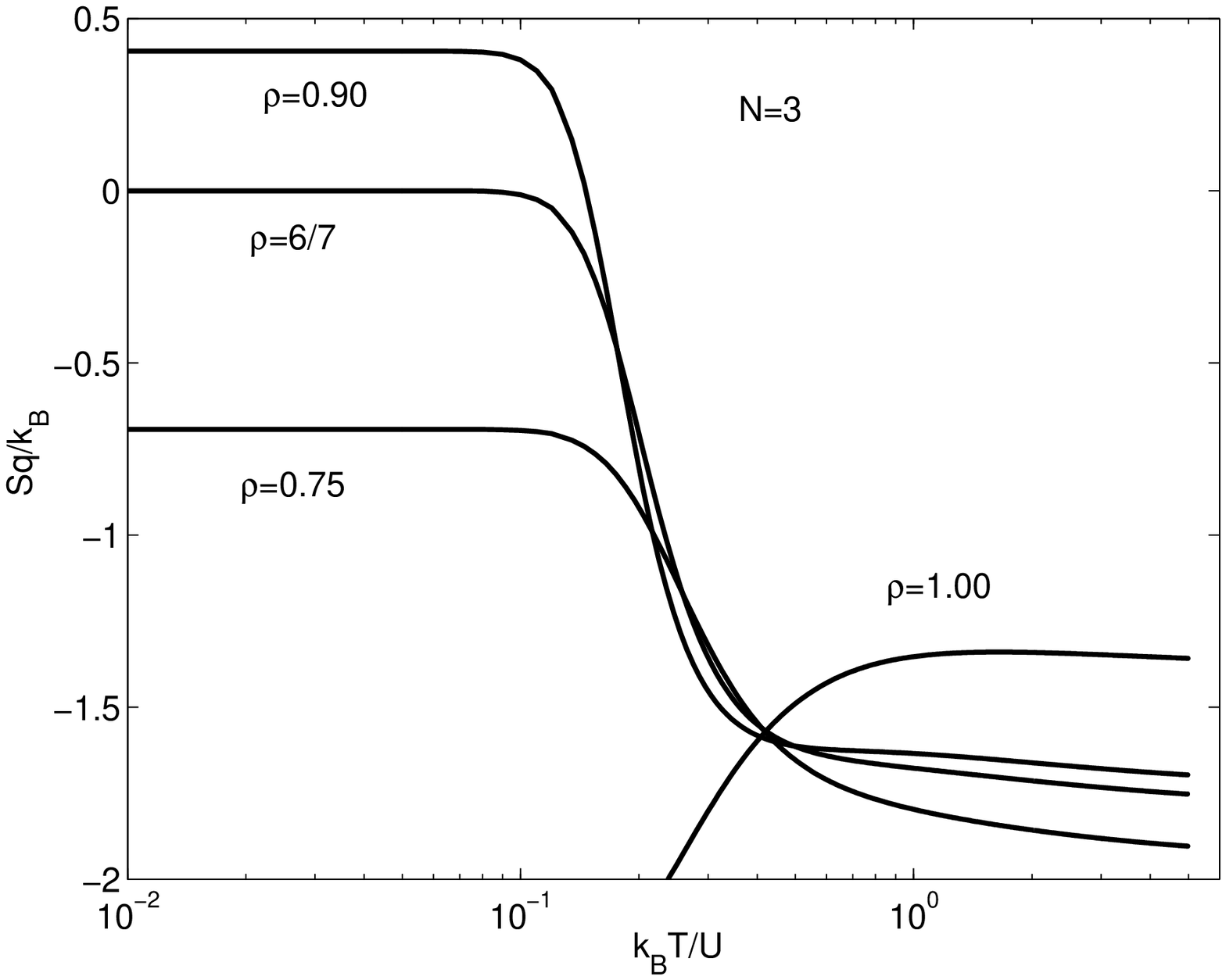} & 
\epsfxsize=3in                                     
\epsfysize=3in
\epsffile{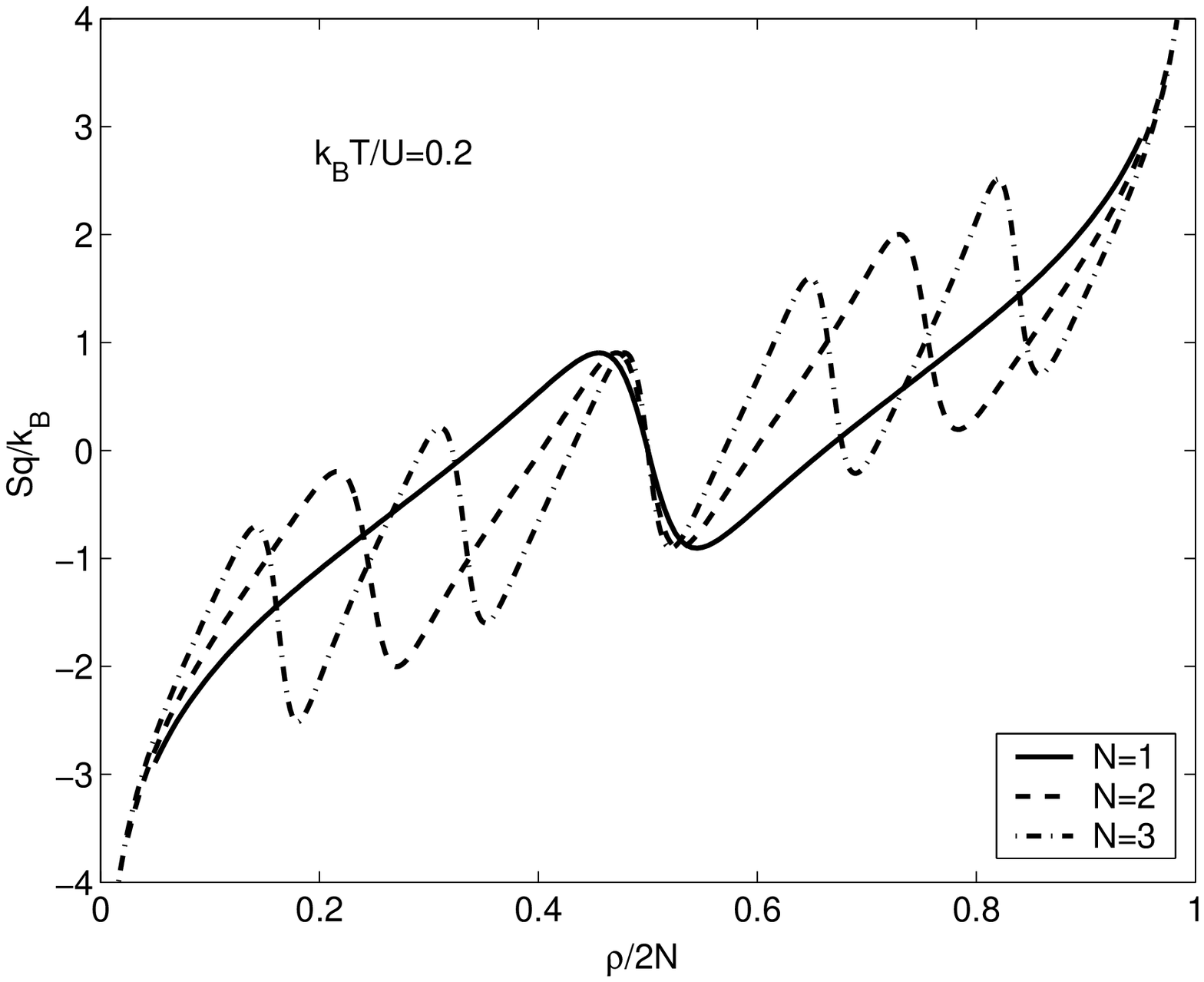} \\ 
\end{array}$
\caption{(Left) Thermopower plotted as a function of $k_BT/U$ for different values of $\rho$ less
than half filling for $N=3$. The thermopower goes to zero in the limit of small $k_BT/U$ for 
$\rho = 6/7$ and is of opposite sign for values of $\rho$ above and below this value. The curve 
for $\rho=1.2$ diverges in this limit. (Right) Thermopower plotted as a function of $\rho/2N$ for 
different values of $N$. The thermopower is non-monotonic and oscillates with the number of maxima 
on either side of half-filling ($\rho/2N=1$) equal to the number of orbitals. These oscillations 
go away in the limit of large $k_BT/U$ when all the curves collapse onto a single curve given by 
Eqn. \ref{gheikes}.}
\end{figure}
\end{widetext}

\section{\label{sec:multib}Multiple orbital Hubbard model in a magnetic field} 
We now consider a multiple orbital Hubbard model in the presence of a magnetic field. Once again, the field is introduced through a 
Zeeman term, which couples differently to different orbitals. The Hamiltonian for such a system is 
\begin{eqnarray}
H & = & -\sum_{j\alpha \beta \sigma}t_{\alpha \beta}c_{j\alpha \sigma}^+c_{j+1\beta \sigma}
+ {\rm h.c.} + U \sum_{j\alpha}n_{j\alpha \uparrow}n_{j\alpha \downarrow} \nonumber \\ & & + \frac{U}{2}\sum_{\stackrel {j\alpha \beta
\sigma \sigma'}{\alpha \neq \beta}}n_{j\alpha \sigma}n_{j\beta \sigma'} + \sum_{j \alpha \sigma}g_{\alpha \sigma}c_{j\alpha 
\sigma}^+c_{j\alpha \sigma} 
\label{multib}
\end{eqnarray}
The charge and heat current operators are given by 
\begin{equation} 
J_e =\frac{qa}{i\hbar}\sum_{j\alpha \beta \sigma}t_{\alpha \beta}c_{j+1\alpha \sigma}^+c_{j\beta \sigma}-t_{\alpha \beta}^{*}c_{j\alpha
\sigma}^+c_{j+1\beta \sigma} 
\label{multibe} 
\end{equation} 
\begin{widetext}
\begin{eqnarray}
J_Q & = & \frac{Ua}{i\hbar}\left[\sum_{j\alpha \beta \sigma}(t_{\alpha \beta}c_{j+1\alpha \sigma}^+c_{j\beta \sigma}
-t_{\alpha \beta}^*c_{j\beta \sigma}^+c_{j+1\alpha \sigma})(Un_{j+1\alpha -\sigma} + g_{\alpha \sigma}B) +
\frac{1}{2}\sum_{\stackrel{j\alpha \beta \gamma \sigma \sigma'}{\gamma \neq \alpha}}(t_{\gamma \beta}c_{j+1\gamma
\sigma'}^+c_{j\beta \sigma'}-t_{\gamma \beta}^*c_{j\beta \sigma'}^+c_{j+1\gamma \sigma'})n_{j+1\alpha \sigma} \right] \nonumber \\
& & + \frac{a}{i\hbar}\sum_{j\alpha \beta \gamma \sigma}t_{\alpha \beta}t_{\beta \gamma}c_{j+2 \gamma \sigma}^+c_{ja \sigma}-t_{\alpha
\beta}^*t_{\beta \gamma}^*c_{j \alpha \sigma}^+c_{j+2 \gamma \sigma}
\label{multibQ}
\end{eqnarray}
\end{widetext}
where the heat current now picks up an additional contribution from the transport of magnetic energy in the same way as Eqn.  
\ref{singQ}. Once again, the thermopower is given by Eqns. \ref{thermo}, \ref{kubo1}, \ref{kubo2} and \ref{trace} where 
\begin{equation} 
I_1 ={\rm Tr} [B(\{n_{1\alpha \sigma_1},n_{2\beta \sigma_2}\})\sum_{\alpha \beta \sigma}|t_{\alpha \beta}|^2(n_{2\alpha \sigma}-n_{1\beta  
\sigma})^2] 
\end{equation} 
and 
\begin{widetext}
\begin{equation} 
I_2 = {\rm Tr} [B(\{n_{1\alpha \sigma_1},n_{2\beta \sigma_2}\})\{\sum_{\alpha \beta \sigma}|t_{\alpha \beta}|^2(n_{2\alpha 
\sigma}-n_{1\beta \sigma})^2 (Un_{2\alpha -\sigma} + g_{\alpha \sigma}B) \nonumber +    
\sum_{\alpha \beta \gamma \sigma \sigma'}|t_{\gamma \beta}|^2(n_{2\gamma \sigma'}-n_{1\beta \sigma'})^2 n_{2\alpha \sigma}\}]
\end{equation} 
after reducing the problem to a trace over two sites. The constraint on the trace is 
\begin{equation} 
U(n_{1\beta-\sigma} + \sum_{\gamma \neq \beta \sigma'}n_{1\gamma \sigma'} - n_{2\alpha -\sigma} - \sum_{\gamma \neq \alpha \sigma'}n_{2\gamma
\sigma'}) + (g_{\beta \sigma} - g_{\alpha \sigma})B = 0 
\label{constr2} 
\end{equation} 
\end{widetext}

This constraint differs from the one in Eqn. \ref{constr1} by the last term involving the magnetic field. This has interesting
consequences as it produces spikes in the thermopower as shown in Fig. 4 at certain integer values of the magnetic field. This can be
understood as follows: The constraint on the trace is just a statement of the conservation of energy. Without a magnetic field, all the
orbitals are degenerate and the conservation of energy condition implies that the number of carriers on the two sites participating in a
hop should be the same because the only energy involved is the Hubbard energy. In the presence of a magnetic field, the degeneracy of
the levels is lifted and now the energy that has to be conserved is a combination of the Hubbard energy and the magnetic energy. The
constraint condition Eqn. \ref{constr2} reduces to Eqn. \ref{constr1} except for special values of the magnetic field when
$(g_{\beta \sigma} - g_{\alpha \sigma})B/U$ is an integer. This involves two different kinds of orbitals ($\alpha$ and $\beta$) where
the Zeeman energy transferred is equal to the negative of the transferred Hubbard energy.  This transfer can take place only through two
different kinds of orbitals because the hops are spin conserving and hence such spikes will not be observed in a system with just one
kind of orbital. This is illustrated in Fig. 3.

The spikes shown in Fig. 4 could be observed in experiments provided the magnitude of the Zeeman coupling is large. This is likely to
happen in compounds with heavy transition and actinide metal ions such as the ruthenates and uranium based compounds where the strongly
correlated electrons involved in transport are in the $f$ orbital. The spikes in an actual experiment will be broadened due to 
thermal effects such as coupling with phonons or some other excitations which will relax the energy conservation constraint. The effect
of phonons on thermopower in strongly correlated systems has been investigated earlier in a different context \cite{barma}.

\begin{figure}[h!]
$\begin{array}{c}
\epsfxsize=3in
\epsfysize=2in
\epsffile{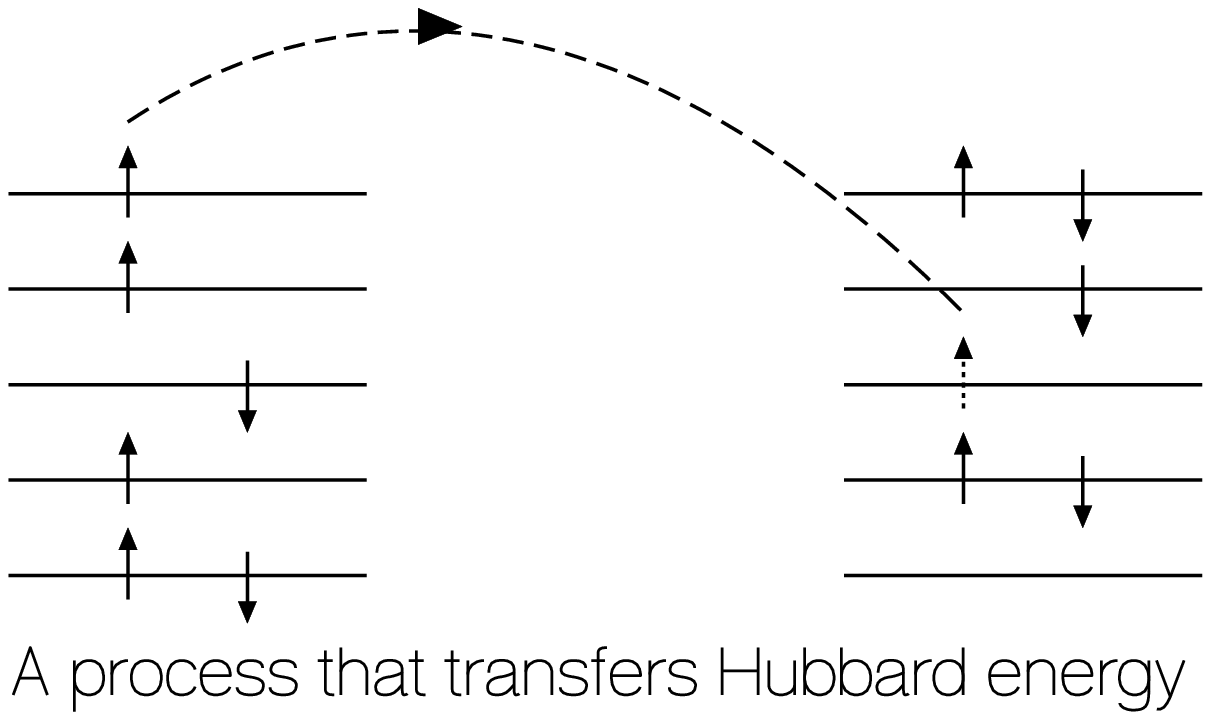} \\
\epsfxsize=3in
\epsfysize=2in
\epsffile{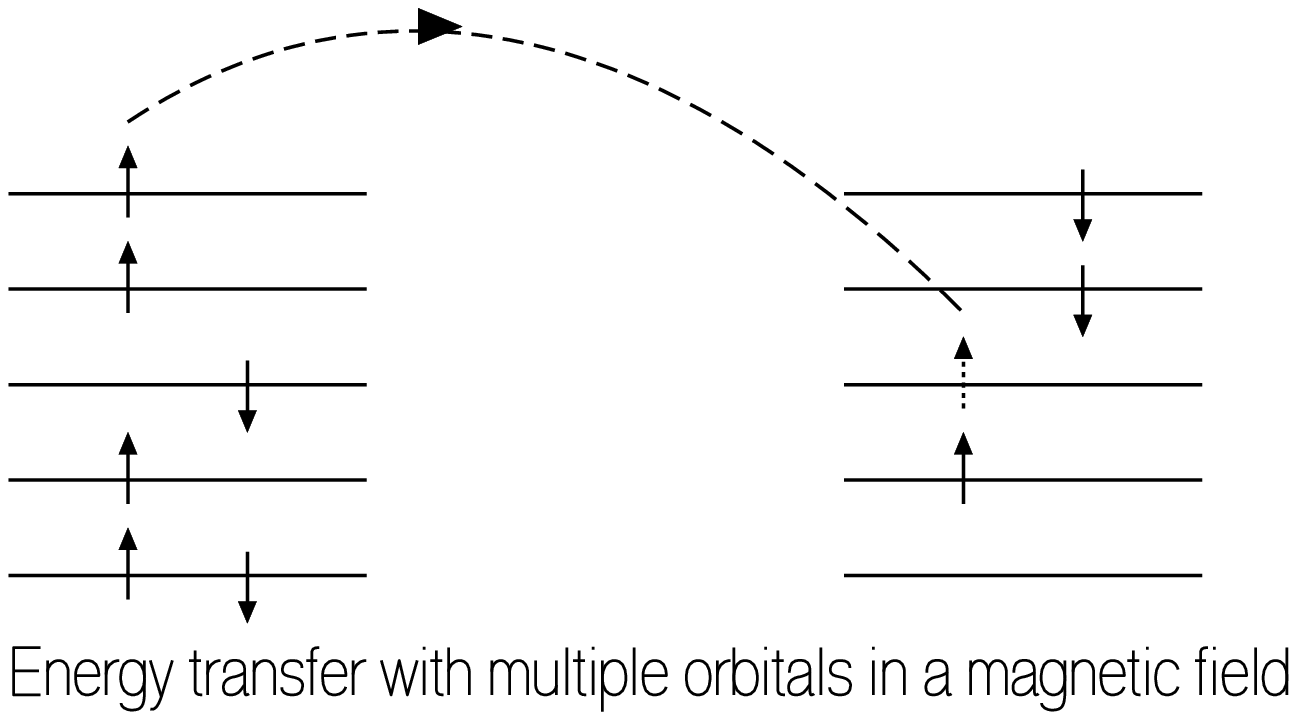} \\
\end{array}$
\caption{(Left) Energy conservation in the absence of a magnetic field. The Hubbard energy has to be conserved and hence the total
number of carriers on the sites participating in the hopping process has to be the same. (Right) Energy conservation with a magnetic     
field. The Hubbard energy and magnetic energy have to be conserved together and thus the number of carriers need not be the
same on the two sites anymore. The difference in Hubbard energy can be compensated by the gain in magnetic energy because of the
different coupling strengths of the orbitals involved in the hop.}
\end{figure}

\begin{figure}[h!]
\label{multimag}
\epsfxsize=3in
\centerline{\epsfbox{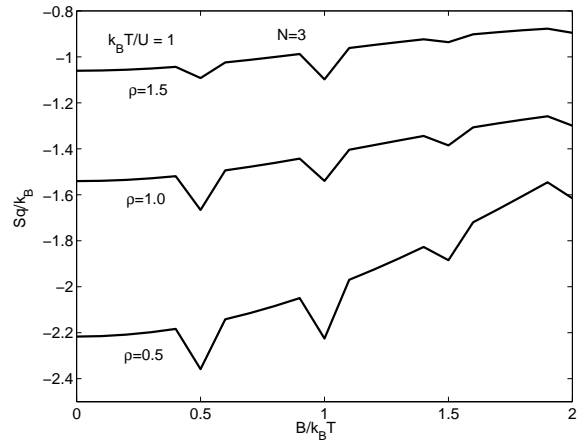}}
\caption{Thermopower as a function of magnetic field for the multiple band Hubbard model with $N=3$. The values of the coupling constant
are $g_{1\downarrow}=-3/2, g_{2\downarrow}=-1/2, g_{3\downarrow}=1/2,g_{1\uparrow}=-1/2, g_{2\uparrow}=1/2, g_{3\uparrow}=3/2$. The
thermopower consequently has spikes as a function of the magnetic field at special values of the field as explained in the text. The
seemingly finite width of the spikes in the figure is due to the discrete nature of the points in the plot.}
\end{figure}  

\section{\label{sec:final}Conclusions and comments} 

We have studied the thermopower in the atomic limit in the presence and absence of a magnetic field in the single and multiple orbital
Hubbard model. We obtained exact expressions for the thermopower and chemical potential and obtained the full temperature dependence of
these quantities between the strong correlation $U/k_B T \gg 1$ and high temperature $U/k_B T \ll 1$ limits where they reduce to the
corresponding Heikes formulae previously obtained using combinatric arguments. The thermopower of the multiple orbital system shows
spikes as a function of magnetic field for certain special values of the magnetic field due to a transfer of commensurate Zeeman
energy along with the Coulomb energy during hopping processes. We conjecture that it might be possible to observe these spikes in some
form in complexes of higher transition metal ions.

\begin{acknowledgments}
The author would like to thank David Huse and Vadim Oganesyan for enlightening discussions and 
reviewing the manuscript.
\end{acknowledgments}

\bibliography{thermo}

\end{document}